\documentclass[twocolumn, letterpaper, pra, showpacs, aps]{revtex4-1}
\usepackage[english]{babel}
\usepackage[applemac]{inputenc}
\usepackage{pgf}
\usepackage{ulem}
\usepackage{amsmath}
\usepackage{amsfonts}
\usepackage{mathtools}
\usepackage{color}
\usepackage{upgreek}
\usepackage{hyperref}
\hypersetup{colorlinks, citecolor=blue, linkcolor=blue, urlcolor=blue}
\newcommand{\ie}{i.~e.,~}

\newcommand{\ket}[1]{\mid #1\rangle}

\definecolor{kentgreen}{RGB}{39, 174, 96}

\newcommand\rmaa{\bgroup\markoverwith{\textcolor{magenta}{\rule[0.5ex]{2pt}{0.4pt}}}\ULon}

\hypersetup{colorlinks, citecolor=blue, linkcolor=black, urlcolor=black}

\begin{document}
\title{Simulating one-dimensional systems with stationary Rydberg dark polaritons}

\date{\today}

\author{Hudson Pimenta}
\email{hpimenta@physics.utoronto.ca}
\author{Aaron Z. Goldberg}
\email{goldberg@physics.utoronto.ca}
\author{Josiah Sinclair}
\author{Kent Bonsma-Fisher}
\affiliation{Department of Physics, University of Toronto, 60
  St. George Street, Toronto, Ontario M5S 1A7, Canada}

\begin{abstract}

Electromagnetically-induced transparency (EIT) with Rydberg atoms enables strong interactions between atoms to be controlled and measured by pulses of light. Consequently, Rydberg dark polaritons are a promising platform for quantum simulations. Of particular interest are simulations of one-dimensional systems, known to exhibit unique properties compared to their higher-dimensional counterparts. One limitation of standard EIT is that reducing the polariton group velocity to bring the dark polaritons to rest also implies activating strong Rydberg interactions. 
To achieve independent control of polariton speed and interaction strength, we propose a stationary light-scheme to bring Rydberg polaritons to a stand-still. This allows them to remain primarily photonic and, therefore, non-interacting, before Rydberg interactions are turned on. 
Once activated, in the dilute regime and in the presence of strong transverse confinement, the strong polarizabilities of Rydberg atoms give rise to an effective contact interaction between polaritons. By tuning the various parameters of the stationary-light scheme, the effective one-dimensional scattering length may be adjusted to yield various kinds of physics, including the regime of a Tonks-Girardeau gas, when polaritons exhibit fermionic behavior. We outline a  protocol to observe the resulting spatial correlations between neighboring polaritons.

\end{abstract}
 
\maketitle

\section{Introduction}
\label{sec:introduction}
There is much interest in probing the dynamics of many-body
quantum-mechanical systems. Finding naturally-occurring systems that
are simple enough to conform to available theoretical models is
not easy, but we now have a plethora of platforms on which many-body
physics can be implemented under controllable conditions \cite{Noh:2017ij}. A few
examples are cold atoms \cite{Bloch:2008gl,Gross:2017do, Bloch:2012jy}, ion traps
\cite{Blatt:2012gr}, and photons interacting via optical nonlinearities \cite{Hartmann:2016er}. Strong optical
nonlinearities can be achieved, for instance, in cavity-QED arrays \cite{Tomadin:2010uo}, semiconductor
microcavities \cite{Carusotto:2013gh}, and Rydberg dark polaritons
\cite{Gorshkov:2011ba}. In this work, we investigate the
use of Rydberg dark polaritons for simulating one-dimensional many-body systems.

A dark polariton is a coherent superposition of a photon
and a collective atomic excitation arising in the context of
electromagnetically induced transparency (EIT) \cite{Fleischhauer:2000bi}. This excitation is created 
when an on-resonant photon is
incident upon a medium of atoms irradiated by a strong, classical coupling beam. Without the strong
coupling beam, the photon would normally be absorbed into the excited
state and dissipated. However, if the coupling beam strongly couples the excited, short-lived state to a
metastable state, the joint photon-atom system evolves, due to
a combination of interference and dissipative effects, to a linear combination of a photon and a $W$-like metastable atomic state --- the dark polariton. 
The dark polariton is adiabatically connected to the photon, but
propagates with a reduced, tunable group velocity, making it a candidate for photon storage and quantum
memory \cite{Fleischhauer:2002eea}.

A Rydberg dark polariton refers to the special case in which the
chosen metastable state is an energy level of a Rydberg atom. Rydberg atoms
have a single valence electron excited to a state with very high principal quantum
number $n$, which comes with various desirable
properties. The state is long-lived, with lifetime proportional to
$n^3$. Also, due to their large polarizabilities, Rydberg atoms
experience strong van der Waals interactions, with strengths
proportional to $n^{11}$. These properties have made Rydberg atoms an
ideal platform for engineering \cite{Lukin:2001bu, Gorshkov:2011ba} and
observing \cite{Urban:2009jd} photon-photon interactions, single-photon sources \cite{Dudin:2012hm}, quantum phase gates with single photons \cite{Tiarks:2016cn}, entanglement generation between individual neutral
atoms \cite{Wilk:2010db}, and other applications in
quantum information \cite{Saffman:2010ky}. Rydberg dark polaritons
inherit the long lifetimes and interaction properties from their atomic
components and, therefore, have also been considered for probing many-body
physics \cite{Bienias:2014bc, Moos:2015fp}.

Of particular interest are one-dimensional many-body systems, partly due their simplicity and partly
because they feature peculiar properties compared to their higher-dimensional counterparts \cite{Cazalilla:2011dma}. With regard to their
simplicity, a few models can be solved exactly \cite{Lieb:1963ik,
  Calogero:1971fw, Sutherland:1971wd}, and there are bosonization schemes \cite{Haldane:1981vf, Cazalilla:2004dd} to extract general
properties of more general models. Regarding their
peculiarities, a few 
examples are the boson-fermion statistics transmutation
(when bosons start behaving like fermions and vice-versa) \cite{Cheon:1999jj} and
spin-charge separation (the decoupling of charge and spin density
waves, which propagate with different speeds in the presence of
interactions) \cite{Giamarchi:2004uc}.

While one-dimensional models were historically introduced as toy
models, there are now various proposals and experiments involving
one-dimensional physics. There have been realizations of the
Tonks-Girardeau gas in the context of cold atoms
\cite{Paredes:2004fp,Kinoshita:2004jp}, and proposals for its
implementation in the context
of optical fibers \cite{Chang:2008fh}. A scheme has also been proposed to
observe spin-charge separation in hollow-core fibers
\cite{Angelakis:2011ji}.
Specifically in the context of Rydberg polaritons, it has been
suggested that the latter could implement the Lieb-Liniger model
\cite{Bienias:2014bc}. The emergence of a Mott insulator in the
presence of a periodic, arbitrarily weak potential has also been
discussed in this context \cite{Otterbach:2013fe} (and previously in the context of cold atoms \cite{Buchler:2003dr, Buchler:2011iq}).
More recently, a scheme
for observing spin-charge separation using two species of Rydberg
polaritons was proposed \cite{Shi:2016jr}.

An important part of simulating a many-body system is also the procedure to create a specific quantum state. In the context of many-body physics, one is often interested in the low-energy domain, particularly in the ground state of a model. Having the effective Hamiltonian of a system match a certain model is one requirement for simulation, but we must also ensure that the preparation protocol leads to a quantum state that remains relatively close to the ground state of the interacting model. In the context of cold atoms, for example, one usually relies on
thermalization. Rydberg polaritons are part of a naturally-driven
system (though dissipation is greatly minimized by the EIT condition),
and so we must make use of another method, such as adiabatic passage.

%There has not been much discussion, however, on how to experimentally approach the ground state of the effective one-dimensional models describing the Rydberg polaritons. Having the effective Hamitonian for polaritons be the same as a model is a requirement, but we must also ensure that the preparation protocol leads to a quantum state that remains relatively close to the ground state of the interacting model, given that many of the theoretical predictions belong to the domain of low-energy physics.

In this work, we investigate a method for adiabatically preparing an
interacting system of one-dimensional Rydberg dark polaritons in a
many-body ground state by using stationary polaritons
\cite{Bajcsy:2003,Zimmer:2008ha}. Stationary polaritons are a variant of dark
polaritons that may be brought to rest while keeping their atomic
component to a minimum, in contrast to the traditional EIT polariton,
which only stops when the photon is stored as an atomic excitation \cite{Zhang:2012fy}. In our
scheme, the first step is to load a light pulse on top of an ensemble
of atoms in a cigar-shaped trap. By having two auxiliary lasers
shining on the atoms,  the initially propagating pulse is converted into
a stationary pulse at rest on top of the atomic ensemble. The
initially small atomic component of the polariton keeps the interaction to a
minimum, this pulse then being relatively close to the ground state of the non-interacting model. Next, the power of the auxiliary lasers is adjusted so as to
increase the atomic component and, consequently, the interaction
strength. This setup follows the spirit of another, made in the
context of hollow-core fibers, which also evokes the concepts of
stationary light and adiabatic passage to approach the ground state of an interacting Hamiltonian
\cite{Chang:2008fh}. Realizing this setup experimentally in the
context of Rydberg polaritons would be an important milestone towards
implementing the more elaborate physics discussed by some of the previous work in
this field. 

 The paper is organized as follows. In Sec.~\ref{sec:1dsystems}, we
 overview one-dimensional systems and some of their more remarkable properties to
 motivate the interest in one-dimensional simulations. In Sec.~\ref{sec:rydberg}, we introduce interacting Rydberg dark polaritons in the context of EIT, and how the interaction properties of Rydberg atoms may be used to implement one-dimensional models. 
In Sec.~\ref{sec:stat-dark-polar}, we review the concept of stationary light, and discuss how Rydberg polaritons combined with stationary light allows us to adiabatically approach the ground state of an interacting one-dimensional system. In Sec.~\ref{sec:preparation}, we discuss the preparation protocol. In Sec.~\ref{sec:conclusions}, we summarize our results and discuss issues that require further examination.

\section{One dimensional systems}
\label{sec:1dsystems}
A one-dimensional system may be seen as the limiting case of a
three-dimensional system in which the transverse degrees of freedom
are frozen due to energy constraints. If the energy gap between the
tranverse modes is large compared to the characteristic energy of
the particles in a system, then only changes in the longitudinal
components are allowed. Such confinement can be achieved with current
technology, using cigar-shaped traps \cite{Burger:1999vs} or hollow-core optical fibers
loaded with atoms \cite{Chang:2008fh}. Under such circumstances, after eliminating the
tranverse modes, one is left with the prototypical Hamiltonian of a one-dimensional system,
\begin{align}
\label{onedimensionalprototype}
  H_{1\text{D}} &= -\frac{1}{2m} \int dx~ \psi^\dagger (x) \frac{\partial^2}{\partial x^2} \psi
  (x) \\
  &+ \int dx \int dx' ~ V(x, x') \rho (x) \rho(x') \nonumber,
\end{align}
where $\psi (x)$ is a bosonic or fermionic field that annihilates particles at position
$x$, $\rho (x) = \psi^\dagger (x) \psi (x)$ is the particle density and $m$ is the particle mass. The first term is the kinetic energy and the second term
describes the interparticle interaction. In this work, we take the field to be bosonic, such that $[ \psi
(x), \psi (x') ] = [ \psi^\dagger (x), \psi^\dagger (x') ] = 0$ and $ [ \psi
(x), \psi^\dagger (x') ] = \delta (x - x')$. Effects of
external potentials may also be accounted for by adding a term of the
form $\displaystyle \int dx~ \mu (x) \rho (x) $ to the
Hamiltonian, where $\mu (x)$ plays the role of the trapping potential. Such is the case, for example, when one wants to quantitatively treat
the effects of an optical trap.

The transverse confinement makes one-dimensional systems
more tractable analytically than their higher-dimensional counterparts,
since it establishes a natural ordering for the particles. More precisely, consider a
one-dimensional system of $N$ particles and their respective
positions $x_1, x_2, ..., x_N$. Since the particles are identical, it
suffices to understand the physics of the ordered subspace $x_1 \le x_2 \le ... \le
x_N$; swapping two particles should make no difference since they are indistinguishable. This greatly simplifies the task of finding
exact solutions.

An exactly solvable model is the Lieb-Liniger model \cite{Lieb:1963ik}, which describes bosons interacting through a
repulsive contact potential $V(x, x') = V_0 \delta (x - x')$, such
that the Hamiltonian of Eq.~\eqref{onedimensionalprototype} takes the
form
\begin{align}
  \label{lieblinigermodel}
  H_{LL} = -\frac{1}{2m} \int dx~ \psi^\dagger (x) \frac{\partial^2}{\partial x^2} \psi
  (x) + V_0 \int dx~ \rho^2 (x).
\end{align}
This model
has been extensively studied in the literature and can be solved exactly using the Bethe-Ansatz approach \cite{Lieb:1963ik}. Originally a toy model
for many-body systems, the Lieb-Liniger model has become particularly relevant with the
development of cold-atom techniques, as it describes the physics of
interacting bosons in cigar-shaped elongated traps and in  the dilute
regime \cite{Lieb:2003kr}. It has also been shown to describe the
physics of two counter-propagating beams in a hollow-core fiber loaded
with atoms \cite{Chang:2008fh}, and of Rydberg dark polaritons under
specific conditions \cite{Bienias:2014bc}. 

To elucidate some of features of this model, here we consider the
somewhat simpler limit of impenetrable bosons, when $V_0 \to +\infty$,
known as a Tonks-Girardeau gas \cite{Girardeau:1960ff}. 

The infinite repulsion leads to the condition that eigenfunctions of the Hamiltonian must vanish when the position of two particles coincide. This requirement is the same as that of a fermionic wavefunction. Thus, the eigenfunctions of the Tonks-Girardeau gas are very similar to that of non-interacting fermions, except that they are still symmetric in their variables.
%The infinite repulsion leads to the condition that the eigenfunctions of the Hamiltonian must vanish when $x_i  = x_j$ for $i \neq j$. This is however, the same requirement as that of a fermionic wavefunction, so we expect the solutions to be proportional to Slater determinants of plane waves, \ie \begin{equation}\phi(x_1, x_2, ..., x_N) \propto \left| \begin{array}{ccc}  e^{i k_1 x_1} & \cdots & e^{i k_1 x_N}\\ \vdots & \ddots & \vdots \\ e^{i k_N x_1} & \cdots & e^{i k_N x_N}  \end{array} \right| \prod\limits_{\substack{i \le N \\ j <i}} \text{sgn} (x_i - x_j),\end{equation} The sign function is included to ensure that the eigenfunction remains symmetric, as required by bosons. 
Due to this similarity, many of the Tonks-Girardeau properties are identical to that of a system of non-interacting
fermions. One important example is the 
Tonks-Girardeau density-density correlation function \cite{Chang:2008fh},
$\displaystyle g_{\text{TG}}^{(2)} (x, x') $, for the ground state:
\begin{align}
\label{correlationdensity}
  g_{\text{TG}}^{(2)} (x, x') 
 =\langle \rho(x) \rho(x') \rangle 
  = 1 - \left[ \frac{\sin k_F (x - x')}{k_F (x - x')} \right]^2,
\end{align}
with $k_F = \pi \rho_0$, where $\rho_0$ is the boson density. The oscillations of $g_{TG}^{(2)} (x, x')$ are a signature of
fermionization, with $k_F$ playing the role of Fermi level, as
it would for a system of $N$ non-interactiong fermions. The physical interpretation behind Eq.~\eqref{correlationdensity} is that, due to the fermionic behavior, when one particle is found at position $x$, we are unlikely to find another near it. Nevertheless,
there are a few differences between a Tonks-Girardeau gas and a system of
non-interacting fermions. Their momentum distribution, for example, is
different \cite{Paredes:2004fp}.

For a general $V(x, x')$, exact solutions are not available, but it is
still possible to make some general predictions for one-dimensional
systems by employing the semi-phenomenological bosonization method based on the harmonic-fluid
approach \cite{Haldane:1981vf, Cazalilla:2004dd}. This approach is predicated on the observation that, since the particles
are transversally confined, the physics should be insensitive to their ordering, only density and phase fluctuations affecting the physical properties of the system. Motivated by this argument, one introduces
a monotonically-increasing field $\theta (x)$ whose value increases by
$\pi$ whenever $x$ exceeds the position of a particle. If the $n$-th particle is
at $x = x_n$, then the field takes the value $\pi n$ at that
point. Hence, around $x_n$, we must have $\displaystyle \delta (x - x_n) = \partial_x \theta
(x) \delta \left [
  \theta (x) - \pi n \right] $, which follows from properties of the delta function. Notice that
$\delta (x - x_n)$ is simply the density around the $n$-th particle,
so this relation allows us to express $\rho (x)$ in terms of $\theta
(x)$. More generally, we have $\displaystyle\rho (x) = \sum_n \delta (x -
x_n) = \sum_n \delta \left[ \theta(x) - \pi n \right] \partial_x
\theta (x)$, which may be rewritten using Poisson's sum formula as
\begin{equation}
  \rho (x) = \frac{\partial_x \theta (x)}{\pi} \sum_{m=-\infty}^{+\infty} e^{i m \theta
    (x) }.
\end{equation}
It is common to measure $\theta(x)$ with respect to the homogeneous
configuration $\pi \rho_0 x$, such that $\theta (x) \to \theta (x) +
\pi \rho_0 x$ in the previous equation:
\begin{equation}
\label{densitybosonized}
  \rho (x) = \left[\rho_0 + \frac{\partial_x \theta (x)   }{\pi} \right] \sum_{m=-\infty}^{+\infty} e^{i m \theta
    (x) + i m \pi \rho_0 x }.
\end{equation}
This is the desired representation for the density in terms of $\theta
(x)$. Similarly, the field $\psi^\dagger (x)$ is represented as
\begin{equation}
\label{fieldbosonized}
  \psi^\dagger (x) \propto \left[ \rho_0 + \frac{\partial_x \theta}{\pi}
  \right]^{1/2} \sum_m e^{i m \theta (x) + i m \pi \rho_0 x } e^{-i \phi
    (x) },
\end{equation}
where we have introduced the phase field $\phi (x)$, whose conjugate
momentum is $\displaystyle\frac{\partial_x \theta(x)}{\pi}$, such that $\left[
  \frac{\partial_x \theta (x) }{ \pi} , \phi (x') \right] = - i \delta (x - x')$. The precise
proportionality factor in Eq.~\eqref{fieldbosonized} factor depends on
the chosen representation of the
delta function. In both Eqs.~\eqref{densitybosonized} and
\eqref{fieldbosonized}, we often only keep the term with $m = 0$, which
captures the long wavelength fluctuations and hence the low-energy
physics.

Represented in terms of $\theta (x)$ and $\phi (x)$, the
harmonic-fluid approach establishes that the effective
low-energy Hamiltonian for the system of
Eq.~\eqref{onedimensionalprototype} takes the form of the so-called
Luttinger liquid model,
\begin{equation}
\label{luttingermodel}
  H_{LL} = \frac{v}{2\pi} \int dx~ \left[ K \left(\partial_x \phi
    \right)^2 + \frac{1}{K}\left( \partial_x \theta \right)^2 \right],
\end{equation}
where $v$ is the speed of propagation of the density waves, and $K$
is the Luttinger parameter. The physical statement of
Eq.~\eqref{luttingermodel} is that the low-energy excitations of a
one-dimensional system are phonons propagating with speed $v$. The
precise values of $v$ and $K$ must be determined either by comparison
with exact solutions, or by evaluating some thermodynamic property,
such as the compressibility.

To understand how different values of $K$ affect the dynamics, it is
useful to look at the density-density correlation function for the ground state of
$H_{LL}$ at large distances:
\begin{align}
  g_\text{TG}^{(2)}\left(x,x^\prime\right) \sim \frac{\cos \left[ 2\pi \rho_0 (x - x') \right]}{ | x - x'|^{2K}}.
\end{align}
When $K$ is larger, the density-density correlation decays faster and the
system behavior resembles a superfluid, whereas lower $K$ makes the
charge-density waves more manifest \cite{Orignac:1998ir}. There
is, however, no true phase transition, but only quasi-long-range order, in
agreement with the Mermin-Wagner theorem \cite{Mermin:1968in}.
The case $K \to +\infty$ corresponds to non-interacting bosons; $K =
1$ corresponds to the Tonks-Girardeau gas, the quadratic decay of $\langle \rho (x) \rho^\dagger (x') \rangle$ consistent with
Eq.~\eqref{correlationdensity}; and, for long-range power-law interactions,
$0 < K \le 1$ \cite{Dalmonte:2010gf}.

The quasi-crystal long-range order corresponding to $K = 0$ is achieved by
introducing an arbitrarily weak periodic potential. The system then
becomes a Mott insulator when the density is commensurate with
the lattice. This is because the potential oscillations cancel some
of the fast oscillations of $\rho (x)$ in
Eq.~\eqref{densitybosonized}. To see this more explicitly, consider
the Hamiltonian associated with an external periodic potential, $H_V =
V_0 \int dx~ \cos (\ell x) \rho (x)$. When $\ell = \pm s \pi \rho_0$, where
$s$ is an integer, it becomes necessary to retain the terms corresponding to $s$ and $-s$ in
Eq.~\eqref{densitybosonized}. Then $H_V$ generates term of the
form $V_0 \int dx~ \cos ( s \theta)$, which, combined with $H_{LL}$,
is also known as the Sine-Gordon model \cite{Coleman:1975tn}. This model is known to describe gapped excitations, which explains the insulating behavior.
This feature has been discussed in the context of cold atoms
\cite{Buchler:2011iq} and also for Rydberg polaritons \cite{Otterbach:2013fe}.

The properties above are fairly general, and grounded solely on the
assumption that the Luttinger liquid picture holds. An even more
striking phenomenon arises when spin is included, leading to two
species of bosons, which we may describe in terms of the total charge
density $\rho_c (x) = \rho_\uparrow (x) + \rho_\downarrow (x)$ and the spin density
$\rho_s (x) = \rho_\uparrow (x) - \rho_\downarrow (x)$. By introducing
the pairs of density and phase fields $\theta_c (x)$ and $\phi_c (x)$,
and $\theta_s(x)$ and $\phi_s(x)$ to describe the charge and spin
 sectors, under special circumstances, the Hamiltonian separates into
 the sum of a Luttinger liquid Hamiltonian for each sector \cite{Angelakis:2011ji}, 
\begin{align}
  H_{sc} = \sum_{\sigma = c, s} \frac{v_\sigma}{2\pi} \int dx~ \left[ K_\sigma \left(\partial_x \phi_\sigma
    \right)^2 + \frac{1}{K_\sigma}\left( \partial_x \theta_\sigma \right)^2 \right],
\end{align}
where $v_c$ and $v_s$ represent the charge and spin wave speeds,
respectively, and $K_c$ and $K_s$ are the Luttinger parameters for the
respective sectors. It is striking that $v_s$ and $v_c$ are in fact
different, meaning that the charge-density and magnetization
fluctuations propagate with different speeds. This phenomenon is known
as the spin-charge separation, and there have been discussions about the
conditions required for observing it with Rydberg polaritons \cite{Shi:2016jr}.

After this brief overview of one-dimensional systems, we now proceed to
discuss Rydberg dark polaritons in the context of EIT and establish
connections with the one-dimensional physics presented here. 

\section{Rydberg polaritons}
\label{sec:rydberg}
\begin{figure}[t]
\includegraphics[width=0.5\columnwidth]{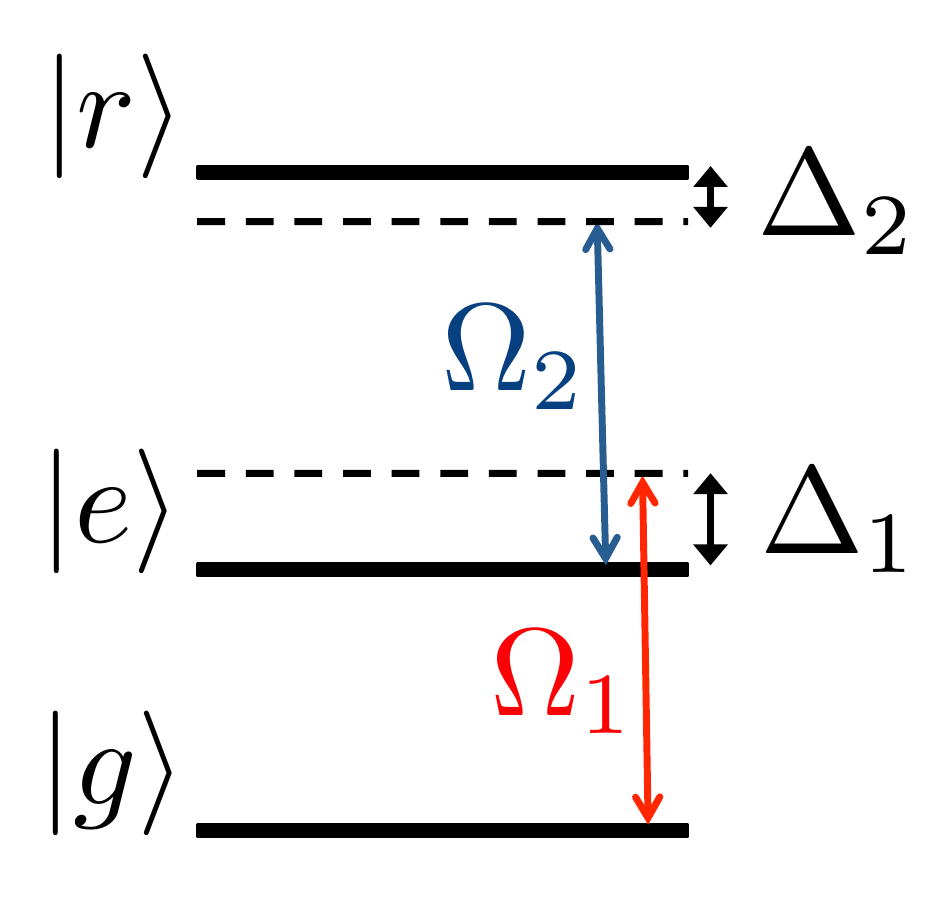}
\caption[3levelscheme]{\label{3levelscheme} EIT ladder scheme. A three-level atom with atomic states $| g
\rangle$, $|e \rangle$, $|r \rangle$ and energies $0$,
$\omega_e$ and $\omega_r$ are
coupled by two lasers of frequency $\omega_1$ and $\omega_2$, and Rabi coupling $\Omega_1$ and $\Omega_2$. The detunings are represented by $\Delta_1$ and $\Delta_2$.
}
\end{figure}

In this section, we recapitulate the physics of dark polaritons in the
context of three-level EIT, and then include the interactions arising from the Rydberg physics. Consider a three-level atom with atomic states $| g
\rangle$, $|e \rangle$, $|r \rangle$ and energies $0$,
$\omega_e$ and $\omega_r$, as illustrated by
Fig.~\ref{3levelscheme}. Levels $|g \rangle$ and $| e \rangle$
($|e\rangle$ and $|r \rangle$) are
coupled by a laser of frequency $\omega_1$
($\omega_2$) and coupling strength $\Omega_1$  ($\Omega_2$). The
Hamiltonian describing this system in the rotating-wave frame of both
lasers is
\begin{align}
H_{\text{EIT}} =&~\Delta_1 \sigma_{ee} + \left( \Delta_1 + \Delta_2 \right)
\sigma_{rr} \\
&+ \Omega_1 \left(\sigma_{eg} + \sigma_{ge} \right) + \Omega_2 \left(
  \sigma_{re} + \sigma_{er} \right), \nonumber
\end{align}
where we have defined $\sigma_{ij} \equiv |i \rangle \langle j |$ (with $i,
j = g, e, r$), and $\Delta_1 = \omega_e - \omega_1$ and $\Delta_2 =
\omega_r - \omega_e - \omega_2$ are the laser detunings with respect
to the atomic transitions they enable. In the EIT setup,
the intermediate state $|e \rangle$ is much shorter-lived than $|g
\rangle$ and $|r \rangle$, which are then assumed to be infinitely
long-lived for simplicity. The state $|r \rangle$ is taken to be a Rydberg state with large principal
quantum number $n$, ensuring a long lifetime. This also
introduces van der Waals interactions, but we consider them
later in this section. $H_{\text{EIT}}$ may be represented in the basis $\{ \ket{g}, \ket{e}, \ket{r} \}$ as the matrix
\begin{align}
\label{Hmatrix}
  H_{\text{EIT}} \to \left[ \begin{array}{ccc} 
0 & \Omega_1 & 0 \\ 
\Omega_1 & \Delta_1 & \Omega_2 \\
0 & \Omega_2 & \Delta_1 + \Delta_2 
\end{array}\right].
 \end{align}.

For simplicity, consider the two-photon resonance case, when $\Delta_2
= -\Delta_1$. In this case, diagonalizing the Hamiltonian yields the
eigenvalues $0$ and $\epsilon_{\pm} = \frac{\Delta_1}{2} \pm \frac{\sqrt{\Delta^2_1 +
    4\Omega^2_1 + 4 \Omega^2_2}}{2}$, with the respective eigenstates
\begin{align}
\label{darkstate}
|D \rangle = \frac{\Omega_2 |g \rangle - \Omega_1 |r\rangle}{\sqrt{\Omega_1^2 + \Omega_2^2}} ;\\
\label{brightplus}
|B_+ \rangle = \frac{ \Omega_1 |g \rangle + \epsilon_+ |e \rangle  +
  \Omega_2 | r \rangle}{\sqrt{\Omega_1^2 + \Omega_2^2 + \epsilon_+^2}} ; \\
\label{brightminus}
|B_- \rangle = \frac{ \Omega_1 |g \rangle + \epsilon_- |e \rangle  +
  \Omega_2 | r \rangle}{\sqrt{\Omega_1^2 + \Omega_2^2 + \epsilon_-^2}}.
\end{align}
The state $| D \rangle $ is called a dark state because it has no population in $| e \rangle$, the only irradiating state on short
timescales. It is, like $|g \rangle$ and $|r \rangle$, 
long-lived. On the other hand, $\ket{B_+}$ and $\ket{B_-}$ are called bright states. An important feature of these states is that, in the
limit $\Omega_2 \gg \Omega_1$, $|D \rangle$ becomes approximately the
ground state $|g \rangle$, while the two remaining bright eigenstates become orthogonal to $|g \rangle$. Hence, by starting with $\Omega_2 \gg \Omega_1$ and slowly tuning the ratio $\Omega_2 / \Omega_1$, an atom initially in state $|g \rangle$ may be made to follow $| D \rangle$ adiabatically. 
In the opposite
limit, $\Omega_2 \ll
\Omega_1$, $|D \rangle$ becomes $- |r \rangle$ (this process is known as stimulated Raman adiabatic passage \cite{Vitanov:2017hn}).

If one is to think of the laser coupling $|g \rangle$ and $|e
\rangle$ as a quantum field, then the conversion of $|g \rangle$ into
$|r \rangle$ really corresponds to storing a photon of this mode,
which can then be retrieved at will through the tuning of $\Omega_1$
and $\Omega_2$. To
store more photons, however, it is necessary to introduce more atoms. Introducing more atoms and treating the laser coupling $|g \rangle$
and $|e \rangle$ as a quantum field with slowly-varying envelope $\mathcal{E} (x)$, the Hamiltonian describing the system is now
\begin{align}
\label{Hamiltonianmanyatoms}
\displaystyle
  H &= \int dx~\mathcal{E}^\dagger (x) \left( - i c \partial_x\right)
  \mathcal{E} (x)  + \Delta \sum_i
  \sigma_{ee}^{(i)}   \\ 
&+ \frac{\tilde{g}}{2} \sum_i \left[ \sigma^{(i)}_{eg} \mathcal{E} (x_i) +
  \sigma^{(i)}_{ge} \mathcal{E}^\dagger (x_i) \right] 
+ \frac{\tilde{\Omega}}{2} \sum_i \left( \sigma^{(i)}_{re} + \sigma^{(i)}_{er} \right) \nonumber.
\end{align}
Here, the index $i$ labels the different atoms. The quantum field is
taken to be centered at the frequency $\tilde{\omega}$ that
realizes a two-photon resonance (hence, the term proportional to
$\sigma^{(i)}_{rr}$ is $0$). The possibility of detuning with regard
to the two-photon resonance is still accounted for by the first term on
the right-hand side of Eq.~\eqref{Hamiltonianmanyatoms}, which corresponds to the
photonic dispersion $\omega_k = c k$, where $c$ is the speed of light and $k$ is the momentum. It is thus implied that frequencies are measured with
respect to $\tilde{\omega}$, so that $k = 0$ really corresponds to the
two-photon resonance condition. The couplings $\Omega_1$ and
$\Omega_2$ have now been renamed $\tilde{g}/2$ and $\tilde{\Omega}/2$,
respectively, and we denote $\Delta_1$ by $\Delta$ at this point. Factors of the form $\displaystyle\exp\left[\pm i
  \left(\frac{\tilde{\omega}}{c}\right) x_i \right]$ arising from the fast
oscillation of the quantum photonic field have already been taken to be
constant and absorbed into $\sigma^{(i)}_{eg}$. Thus,
Eq.~\eqref{Hamiltonianmanyatoms} assumes that the atoms are not
moving, \ie a cold gas. We have also assumed transverse confinement and ignored transverse degrees of freedom. A proper discussion of the elimination of transverse degrees of freedom can be found elsewhere \cite{Moos:2015fp,Bienias:2016bw}.

Next, we introduce the fields $\displaystyle\sigma_{kl} (x) \equiv \sum_i \sigma^{(i)}_{kl}
\delta (x - x_i)$ (with $k, l = g,e,r$) to rewrite the Hamiltonian as
\begin{align}
    H = \int dx~\mathcal{E}^\dagger (x) \left( - i c \partial_x\right)
  \mathcal{E} (x) 
+ \Delta \int dx~ \sigma_{ee} (x)  \nonumber \\
+ \frac{\tilde{g}}{2} \int dx \left[ \sigma_{eg} (x) \mathcal{E} (x) +
  \sigma_{ge} (x) \mathcal{E}^\dagger (x) \right]
\nonumber \\
+ \frac{\tilde{\Omega}}{2} \int dx~ \left[ \sigma_{re} (x) + \sigma_{er} (x) \right].
\end{align}
Based on the Holstein-Primakoff  transformation \cite{Holstein:1940kr}, we introduce bosonic
fields $\mathcal{P} (x)$ and $\mathcal{S} (x)$ such that
$\sigma_{eg} (x)/2 \approx \sqrt{\rho(x) } \mathcal{P}^\dagger (x)$,
$\sigma_{rg} (x)/2 \approx \sqrt{\rho(x) } \mathcal{S}^\dagger (x)$, and
$\sigma_{ee} (x) \approx \mathcal{P}^\dagger (x) \mathcal{P} (x)$, where
$\rho(x)$ is the atomic density. The
Hamiltonian then takes the form 
\begin{align}
\label{Hamiltonianintermsoffields}
  H = \int dx~\left[ \begin{array}{c} \mathcal{E} (x)
               \\\mathcal{P} (x) \\
  \mathcal{S} (x) \end{array} \right]^\dagger
\left[ \begin{array}{ccc} 
-i c \partial_x & g & 0 \\ 
g & \Delta & \Omega \\
0 & \Omega & 0 
\end{array}\right] 
\left[ \begin{array}{c} \mathcal{E} (x) \\
               \mathcal{P} (x) \\ \mathcal{S} (x) \end{array}\right],
\end{align}
where we have absorbed any outstanding constant factors into the couplings $g$ and
$\Omega$, since we are mostly interested in the form of Eq.~\eqref{Hamiltonianintermsoffields}. For simplicity, we assume homogeneous density. The mapping from Pauli to bosonic
operators as used here remains valid in the weak-probing limit, away from saturation. 

Going to momentum space amounts to replacing $-i c \partial_x$ by $c
k$, and the operators by their Fourier transforms, in which case the matrix connecting the creation and annihilation operators
takes the same form as the matrix in Eq.~\eqref{Hmatrix}. Hence, for $k \approx 0$
(two-photon resonance), the eigenvectors are isomorphic to the ones
listed in Eqs.~\eqref{darkstate}, \eqref{brightplus} and
\eqref{brightminus}:
\begin{align}
\label{darkstateop}
\psi_D (x) = \frac{\Omega \mathcal{E} (x) - g \mathcal{S} (x)}{\sqrt{\Omega^2 + g^2}} ;\\
\label{brightplusop}
\psi_+ (x) = \frac{g \mathcal{E} (x) + \epsilon_+ \mathcal{P} (x)  +
  \Omega \mathcal{S} (x) }{\sqrt{\Omega^2 + g^2 +\epsilon_+^2}}; \\
\label{brightminusop}
\psi_- (x) = \frac{g \mathcal{E} (x) + \epsilon_- \mathcal{P} (x)  +
  \Omega \mathcal{S} (x)}{\sqrt{\Omega^2 + g^2 + \epsilon_-^2}},
\end{align}
with $\epsilon_{\pm} = \frac{\Delta}{2} \pm \frac{\sqrt{\Delta^2 +
    4 g^2 + 4 \Omega^2}}{2}$.
Strictly speaking, the relations above are exact only for the Fourier
component of the fields with $k = 0$. That said, we assume that the photons
associated with $\mathcal{E} (x)$ are sufficiently narrow in $k$ for
these relations to approximately hold.

The field $\psi^\dagger_D (x)$ is analogous to the dark state of
Eq.~\eqref{darkstate}, and creates a boson consisting of a linear combination of a photon
and a collective atomic excitation around position $x$. We call this excitation the dark polariton. The excitations created by $\psi^\dagger_{+} (x)$ and $\psi^\dagger_- (x)$ are the bright polaritons, which are much shorter-lived, just like the states $\ket{B_+}$ and $\ket{B_-}$. 

\begin{figure}[h]
\includegraphics[width=1.1\columnwidth]{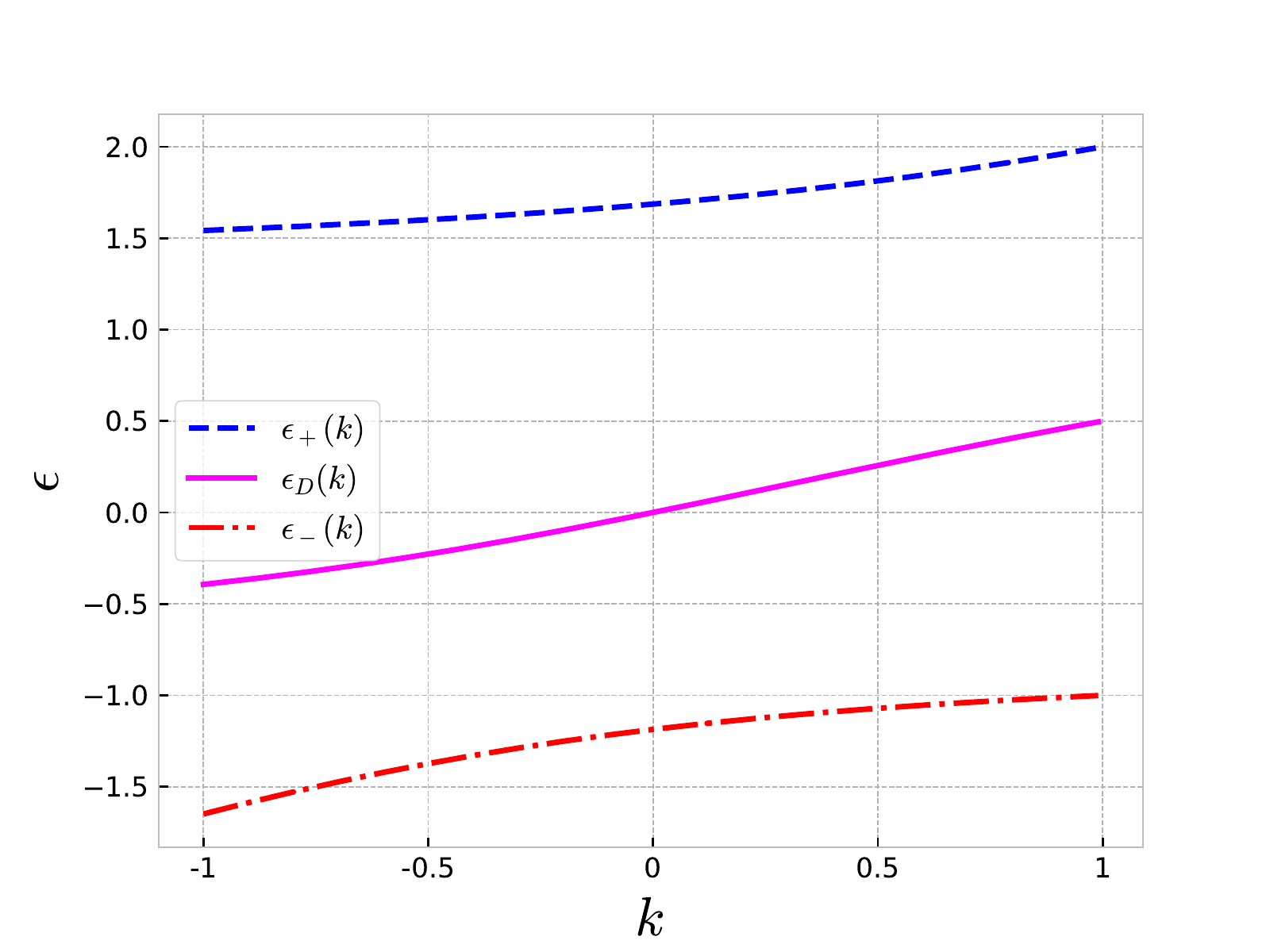}
\caption[darkpoldisp]{\label{darkpoldisp} EIT polaritons dispersion. The intermediate branch depicts the dark polariton dispersion, while the upper and lower branches correspond to the bright modes.}
\end{figure}

The dispersion for each polariton branch is shown in Fig.~\ref{darkpoldisp}. The intermediate branch corresponds to the dark polariton. Its speed of propagation and effective mass can be determined by looking at the dispersion around $k = 0$ \cite{Bienias:2016gs}:
\begin{align}
  \epsilon_D (k) \approx v k + \frac{k^2}{2m}, \\
\label{speedpolariton}
v = \frac{\Omega^2}{\Omega^2 + g^2} c ,\\
\label{effectivemass}
m = \frac{\left(g^2 + \Omega^2\right)^3}{2c^2 g^2 \Omega^2 \Delta}.
\end{align}
Eq.~\eqref{speedpolariton} shows that the dark polariton speed can be controlled by adjusting $\Omega$ relative to $g$. When $\Omega \ll g$, the excitation becomes mostly atomic and the speed approaches $0$. 

We also notice that, like $| D \rangle$ and $| g \rangle$ in
Eq.~\eqref{darkstate}, $\psi^\dagger_D (x)$ and $\mathcal{E}^\dagger (x)$ are
adiabatically connected, so that any photonic state $f [\mathcal{E}^\dagger
(x) ] | 0 \rangle$, where $f(x)$ is an arbitrary function, can be transformed
into the polaritonic state $f [\psi^\dagger_D
(x) ] | 0 \rangle$. The possiblity of controlling the polariton speed by
adjusting the ratio $\Omega/g$, and the polariton long lifetime [owing
to $\psi_D (x)$ having no $\mathcal{P} (x)$ component], historically led to interest in the dark polariton as a candidate for
photon storage and quantum memory \cite{Fleischhauer:2002ee}.

The dark polaritons can be made to interact by letting $|r \rangle$ be a Rydberg atomic level with high principal quantum number $n$. The resulting van der Waals interaction between atoms in state $\ket{r}$ is modeled by
\begin{equation}
H_I = \sum_{i \neq j} \frac{\tilde{C}_6}{ | \boldsymbol{x}_i -
      \boldsymbol{x}_j |^6 } \sigma^i_{rr}\otimes  \sigma^j_{rr}.
\end{equation}
The mechanisms behind this interaction have been previously discussed
in the literature \cite{Browaeys:2016fb}, and the scaling with distance has been verified
experimentally \cite{Beguin:2013dk}.

In terms of the field operators previously introduced, the interaction Hamiltonian becomes
\begin{equation}
\label{InteractionH}
  H_I =  \int dx \int dx' \frac{C_6}{ |x - x' |^6 }  \rho_\mathcal{S} (x) \rho_\mathcal{S} (x'),
\end{equation}
where $\rho_\mathcal{S} (x) = \mathcal{S}^\dagger (x) \mathcal{S}
(x)$. This term leads to interactions
between dark polaritons \cite{Bienias:2016bw, Moos:2015fp}. This can be seen by inverting Eqs.~\eqref{darkstateop},
\eqref{brightplusop} and \eqref{brightminusop} for the dark and bright
modes, which, in particular, yields
\begin{align}
\label{spinwaveoperator}
  \mathcal{S} (x) = - \frac{g}{\sqrt{g^2 + \Omega^2}} \psi_D (x) \\
    + \frac{\Omega}{\sqrt{g^2 +\Omega^2 + \epsilon_{+}^2}} \psi_+ (x)
  \nonumber \\
+ \frac{\Omega}{\sqrt{g^2 +\Omega^2 + \epsilon_{-}^2}} \psi_- (x)
  \nonumber .
\end{align}
To understand the content of Eq.~\eqref{spinwaveoperator}, let us
consider the regime $\Delta \gg \Omega \gg g$, in which case
$\epsilon_+ \approx \Delta$, $\epsilon_- \approx 0$ and
\begin{align}
\label{spinwaveoperatordispersive}
  \mathcal{S} (x) \approx - \frac{g}{\Omega} \psi_D (x)   + \frac{\Omega}{\Delta} \psi_+ (x)
+  \psi_- (x).
\end{align}
We see then that $S (x)$ is weighed mostly by the lower polariton
$\psi_- (x)$, the upper polariton contribution being suppressed by
$\Delta$. $S(x)$ also contains a bit of $\psi_D (x)$, albeit
normalized by the factor $- g/\Omega$. Writing the interaction
Hamiltonian of Eq.~\eqref{InteractionH} in terms of the
polaritonic operators leads to interaction between dark
polaritons, i.e., $\rho_\mathcal{S} (x) \rho_\mathcal{S} (x') \to
\rho_D (x) \rho_D (x')$, the interaction strength normalized by the
factor $g^4 / \Omega^4$:
\begin{equation}
\label{interactingweakdark}
    H_I =  \frac{g^4}{\Omega^4} \int dx \int dx' \frac{C_6}{ |x - x' |^6 }  \rho_\mathcal{D} (x) \rho_\mathcal{D} (x').
\end{equation}

Technically, scattering to bright modes is also possible. To suppress this process, we must ensure that the energy gap between the dark polariton and the bright polaritons is large compared to the typical interaction strength. For large $\Delta$, the gap between dark and upper branches are of the order of $\Delta$, while the gap between dark and lower branches is of the order of $\displaystyle\frac{g^2 + \Omega^2}{\Delta}$. It is important to ensure that the typical interaction strength remains somewhat smaller than these gaps at all times to minimize dissipation. 

It has been suggested that, in the dilute regime, we can replace the van der Waals interaction by an
effective contact interaction, as is often done for cold atoms \cite{Bienias:2014bc},
\begin{equation}
  \frac{C_6}{|x - x'|^6} \to V_0 \delta (x - x'),
\end{equation}
where $\displaystyle V_0 = - \frac{1}{m a_{1\text{D}}}$, and
$a_{1\text{D}}$ is the one-dimensional scattering length. The interaction
Hamiltonian then becomes
\begin{equation}
\label{effectivecontactHamiltonian}
  H_I =  V_0  \int dx~ \rho_{D} (x) \rho_{D} (x).
\end{equation}
Once interbranch scattering has also been neglected, the Hamiltonian for the dark polaritons reduces to
\begin{align}
\label{Hdarkplusint}
  H_{\text{DP}} = \int dx~ \psi^\dagger_D (x) \left( -i v \partial_x -
    \frac{\partial_x^2}{2m} \right) \psi_D (x) \\ \nonumber
+ V_0 \int dx~ \rho_D
  (x) \rho_D (x), 
\end{align}
with $v$ and $m$ previously given in Eqs.~\eqref{speedpolariton} and \eqref{effectivemass}, respectively.

When $v = 0$ and $V_0 > 0$, the Hamiltonian in Eq.~\eqref{Hdarkplusint} corresponds
to the Lieb-Liniger model introduced in Eq.~\eqref{lieblinigermodel}. Thus, this suggests the possibility of
implementing some of the one-dimensional physics discussed in Sec.~\ref{sec:1dsystems} through interacting dark
polaritons. That said, for the dark polaritons described here,
obtaining $v = 0$ requires that we reach the regime $\Omega \ll g$,
which, however, also affects the interaction strength. 
It would be
desirable to have a way to control these parameters
independently. 

Moreover, we would like to have a method such that a sample of
non-interacting dark polaritons may be prepared in the
ground state of the effective non-interacting model, and then the
interaction strength adiabatically turned on, so as to make the state
follow the interacting ground state. This would allow us to observe
the typical signatures of one-dimensional systems discussed in
Sec.~\ref{sec:1dsystems}.
It turns out that a scheme for preparing stationary polaritons
available the literature \cite{Zimmer:2008ha} provides us with what we need. We review this scheme in the next section and adapt it to the context of Rydberg polaritons. 

\section{Stationary dark polaritons}
\label{sec:stat-dark-polar}
The dark polaritons discussed in the previous section hold potential
for implementing interesting physics, but their propagation speed and
interaction strength cannot be independently tunable. By working with stationary polaritons, however, we can control the
polaritons' speed and, in fact, stop light, while still keeping the atomic
component and, hence, the interaction strength, to a minimum. In doing so, we can approach the ground state of the non-interacting effective model, and subsequently tune the interaction strength adiabatically to approach the interacting ground state. 

\begin{figure}[t]
\includegraphics[width=\columnwidth]{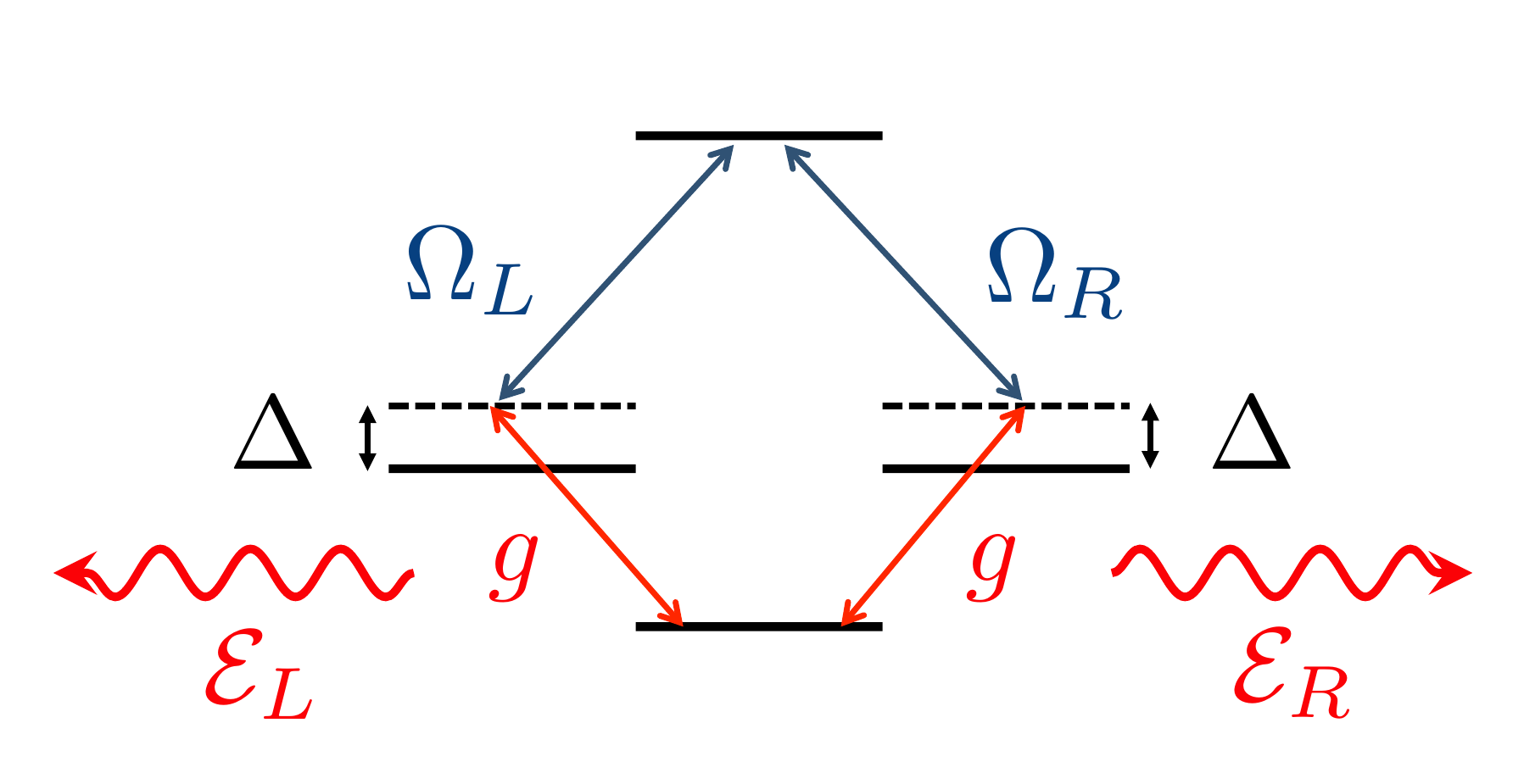}
\caption[stationaryscheme]{\label{stationaryscheme} Stationary-light scheme. The ground state $|g \rangle$, at the bottom of the diagram, is coupled to intermediate states $|e_1 \rangle$ and $|e_2 \rangle$ by quantum pulses $\mathcal{E}_L$ and $\mathcal{E}_R$, respectively (state labels have been omitted to simplify the diagram). The intermediate states are connected to the upper state $|r \rangle$ through classical fields with Rabi coupling $\Omega_L$ and $\Omega_R$. 
}
\end{figure}

Originally proposed in the context of hot vapors \cite{Bajcsy:2003},
stationary polaritons have been extended to cold gases \cite{Zimmer:2008ha,Iakoupov:2016iu}. The
level-scheme in which we are interested is illustrated in Fig.~\ref{stationaryscheme}.
The (lower) ground state $| g \rangle$ is coupled through two two-photon-resonance channels to the upper state $| r \rangle$ via intermediate states $|e_1 \rangle$ and $|e_2 \rangle$. The collective atomic excitations associated with $|e_1 \rangle$, $|e_2 \rangle$ and $|r \rangle$ are described by the fields $\mathcal{P}_L$, $\mathcal{P}_R$ and $\mathcal{S}$, respectively. State $|g \rangle$ is connected to $|e_1 \rangle$ and $| e_2 \rangle$ through quantum fields $\mathcal{E}_L$ and $\mathcal{E}_R$, which propagate to the left and to the right, respectively. We will see soon that the condition of counter-propagation is necessary for a stationary pattern to emerge. 

Mathematically, the time-independent Hamiltonian underlying this scheme is the straightforward extension of the Hamiltonian we derived in Eq.~\eqref{Hamiltonianintermsoffields}:
  \begin{align}
\label{Hamiltonianintermsoffieldsstationary}
 H_S = \int dx\left[ \begin{array}{c} \mathcal{E}_R  \\ \mathcal{E}_L \\
\mathcal{S} \\ \mathcal{P}_R  \\\mathcal{P}_L  \end{array} \right]^\dagger
\left[ \begin{array}{ccccc} 
-i c \partial_x & 0 & 0 & g & 0 \\ 
0 & i c \partial_x  & 0 & 0 & g \\
0 & 0 & 0 & \Omega_R & \Omega_L \\
g & 0 & \Omega_R & \Delta_R & 0 \\
0 & g & \Omega_L & 0 & \Delta_L \\ 
\end{array}\right] 
\left[ \begin{array}{c} \mathcal{E}_R \\ \mathcal{E}_L \\ \mathcal{S} \\
               \mathcal{P}_R \\ \mathcal{P}_L \end{array} \right].
\end{align}
We assume for simplicity that $\Delta_R = \Delta_L = \Delta$.

Working in momentum space, we replace $-i c \partial_x$ by $ck$,
and the fields by their Fourier transforms. We then focus on $k = 0$,
when, like in the standard EIT case, expressions for the eigenvectors
and eigenvalues are relatively simple. The $R$ and $L$ swapping symmetry of
the Hamiltonian can be exploited by introducing the symmetric and anti-symmetric (under $R \leftrightarrow L$) combinations of fields
\begin{align}
  \mathcal{E}_{+} = \frac{\Omega_R}{\Omega} \mathcal{E}_R +
  \frac{\Omega_L}{\Omega} \mathcal{E}_L, \\
\mathcal{E}_{-} = \frac{\Omega_L}{\Omega} \mathcal{E}_R -
  \frac{\Omega_R}{\Omega} \mathcal{E}_L, \\
\mathcal{P}_{+} = \frac{\Omega_R}{\Omega} \mathcal{P}_R +
  \frac{\Omega_L}{\Omega} \mathcal{P}_L, \\
\mathcal{P}_{-} = \frac{\Omega_L}{\Omega} \mathcal{P}_R -
  \frac{\Omega_R}{\Omega} \mathcal{P}_L, 
\end{align}
where $\Omega = \sqrt{\Omega_R^2 + \Omega_L^2}$. Then it turns out
that, at least around $k = 0$, the symmetric fields decouple from
the anti-symmetric ones.

Indeed, it may be seen from Eq.~\eqref{Hamiltonianintermsoffieldsstationary} that $\mathcal{S}$ 
couples to $\mathcal{E}_+$, but not to $\mathcal{E}_-$. The term $\Delta \left( \mathcal{E}^\dagger_R
\mathcal{E}_R + \mathcal{E}^\dagger_L \mathcal{E}_L \right)$,
expressed in terms of symmetric and anti-symmetric fields, yields
simply  $\Delta \left( \mathcal{E}^\dagger_+
\mathcal{E}_+ + \mathcal{E}^\dagger_- \mathcal{E}_- \right)$, i.e., it
does not mix symmetric and anti-symmetric parts. The same is true of
$g\left( \mathcal{P}^\dagger_R \mathcal{E}_R + \mathcal{P}^\dagger_L
  \mathcal{E}_L + \text{H. c.}\right)$, which contributes with $g \left( \mathcal{P}^\dagger_+
  \mathcal{E}_+ + \mathcal{P}^\dagger_-
  \mathcal{E}_- + \text{H. c.} \right)$. It is in fact the kinetic
energy term that leads to some coupling, since
\begin{align}
  \mathcal{E}^\dagger_R ( - i c \partial_x) \mathcal{E}_R + \mathcal{E}^\dagger_L
  ( i c \partial_x) \mathcal{E}_L  = \nonumber \\
  \left[ \begin{array}{c} 
\mathcal{E}_+  \\
\mathcal{E}_- 
\end{array}\right]^\dagger
  \left[ \begin{array}{cc} 
-i u \partial_x & - 2 i c \frac{\Omega_R \Omega_L}{\Omega^2} \partial_x   \\
- 2 i c\frac{\Omega_R\Omega_L}{\Omega^2} \partial_x &  i u \partial_x 
\end{array}\right]
  \left[ \begin{array}{c} 
\mathcal{E}_+ \\
\mathcal{E}_- 
\end{array}\right],
\end{align}
where $\displaystyle u =  c \frac{\Omega^2_R - \Omega_L^2}{\Omega^2}$ plays the role of speed for the symmetric field (the anti-symmetric field propagates with the same speed, but in the opposite direction). The off-diagonal elements couple $\mathcal{E}_+$ and
$\mathcal{E}_-$, but the coupling vanishes for $k = 0$, so we ignore
it for now (we will revisit this problem in the conclusion). Then $\mathcal{E}_+$, $\mathcal{P}_+$ and $\mathcal{S}$
form their own sector, whose Hamiltonian is 
\begin{align}
  H_{+} = \int dx\left[ \begin{array}{c} \mathcal{E}_+ (x)
               \\\mathcal{P}_+ (x) \\
  \mathcal{S} (x) \end{array} \right]^\dagger
\left[ \begin{array}{ccc} 
-i v \partial_x & g & 0 \\ 
g & \Delta & \Omega \\
0 & \Omega & 0 
\end{array}\right] 
\left[ \begin{array}{c} \mathcal{E}_+ (x) \\
               \mathcal{P}_+ (x) \\ \mathcal{S} (x) \end{array}\right].
\end{align}
This Hamiltonian is formally identical to that of Eq.~\eqref{Hamiltonianintermsoffields}, with $c$ replaced by
$u$. Hence, its eigenstates around $k \approx 0$ are formally
identical to those of that Hamiltonian, with $\mathcal{E} (x)$ and
$\mathcal{P} (x)$ replaced by $\mathcal{E}_+ (x)$ and $\mathcal{P}_+
(x)$, respectively. There is still a dark polariton, $\psi_D (x) \sim \Omega
\mathcal{E}_+ (x) - g \mathcal{S} (x)$, i.e., with no components in
$\mathcal{P}_+ (x)$ [nor $\mathcal{P}_- (x)$], adiabatically connected to
$\mathcal{E}_+ (x)$, and moving with speed $u_D = \frac{\Omega^2}{\Omega^2 +
  g^2} u$, (notice again $u$ playing the role of $c$). The speed $u_D$ may be further simplified and written as 
\begin{align}
\label{speedofstationarypolariton}
  u_D = \frac{\Omega^2_R - \Omega^2_L}{\Omega_L^2 + \Omega_R^2 + g^2} c.
\end{align}
Eq.~\eqref{speedofstationarypolariton} opens up the possibility of
controlling the polariton speed independently of the interaction
strength. By working in the limit $g \ll \Omega$, it is still possible
to stop the polaritons, by transitioning from the classical EIT regime
where $\Omega_R \gg \Omega_L$ to the stationarity regime where $\Omega_R
= \Omega_L$ and $u_D = 0$, while keeping the atomic component and, consequently, the interaction strength to a minimum. The relative sign leading to the cancelation in Eq.~\eqref{speedofstationarypolariton} can be traced back to the different slopes for the dispersions of $\mathcal{E}_L$ and $\mathcal{E}_R$.

The polariton mass can also be determined by looking at its dispersion
relation \cite{Zimmer:2008ha}. In the limit $\Omega_R \gg \Omega_L$, it is given by Eq.~\eqref{effectivemass}, with $\Omega_R$ playing the role of $\Omega$. The other relevant limit is $\Omega_R = \Omega_L$, when stationarity has been achieved. In this limit, 
\begin{align}
m = \frac{g^2 \left( g^2 + \Omega^2 \right)}{2 c^2 \Omega^2 \Delta}.
\end{align}

Once a stationary ensemble of non-interacting dark polaritons
has been created on top of the atomic cloud, decreasing $\Omega$
(while still mantaining the condition $\Omega_R = \Omega_L$) increases
the atomic component of the dark polariton and, thus, the
interaction strength. In the weakly-interacting limit, the interaction
Hamiltonian is the same as the one in
Eq.~\eqref{interactingweakdark}. Then we consider the dilute regime
to rewrite the effective interaction Hamiltonian in terms
of an effective contact interaction, as in
Eq.~\eqref{effectivecontactHamiltonian}. Finally, the Hamiltonian for
the dark polaritons takes the form of that of Eq.~\eqref{Hdarkplusint},
but with the speed set to $0$:
\begin{align}
\label{liebliniger}
  H_D = - \frac{1}{2m} \int dx~ \psi^\dagger_D (x)   \frac{\partial^2 }{\partial x^2} \psi_D
  (x) \\ \nonumber
+ V_0 \int dx~ \rho_D
  (x) \rho_D (x). 
\end{align}

Certain phase matching conditions must be satisfied in order to achieve stationarity in the context of Rydberg atoms. The derivation of Eq.~\eqref{Hamiltonianintermsoffieldsstationary} starts from a time-dependent Hamiltonian. Eliminating the time-dependence requires a specific match of the frequencies and momenta of the beams associated with $\mathcal{E}_L$, $\mathcal{E}_R$, $\Omega_L$ and $\Omega_R$. Let the average momenta of each of these beams be, respectively, $\mathbf{k}_{1L}$, $\mathbf{k}_{1R}$, $\mathbf{k}_{2L}$, and $\mathbf{k}_{2R}$. The respective frequencies are $\omega_{1L} = \lvert \mathbf{k}_{1L}\rvert$, $\omega_{1R} = \lvert \mathbf{k}_{1R}\rvert$, $\omega_{2L} = \lvert \mathbf{k}_{2L}\rvert$, and $\omega_{2R} = \lvert \mathbf{k}_{2R}\rvert$ (we have taken $c = 1$). Then the phase-matching conditions leading to the Hamiltonian of Eq.~\eqref{Hamiltonianintermsoffieldsstationary} are
\begin{align}
\mathbf{k}_{1L} + \mathbf{k}_{2L} = \mathbf{k}_{1R}  + \mathbf{k}_{2R}, \\
\label{matchingoftheenergies}
\lvert \mathbf{k}_{1L}\rvert +   \lvert \mathbf{k}_{2L}\rvert = \lvert \mathbf{k}_{1R}\rvert + \lvert \mathbf{k}_{2R}\rvert .
\end{align}

The phase matching conditions are too restrictive for a collinear scheme. A collinear scheme would require the energy spacing between $|g \rangle$ and $|e_i \rangle$ to be the same as that of $|e_i \rangle$ and $|r \rangle$ (for $i = 1,2$). This condition cannot even approximately be satisfied by the typical species used to implement Rydberg atoms. A solution is to keep $\mathcal{E}_L$ and $\mathcal{E}_R$ collinear, and use a coplanar scheme for $\Omega_L$ and $\Omega_R$; we can account for the asymmetric energy spacing by using the components of $\mathbf{k}_{2L}$ and $\mathbf{k}_{2R}$ perpendicular to the direction of propagation of $\mathcal{E}_L$ and $\mathcal{E}_R$. 

Let us denote the direction of propagation of $\mathcal{E}_L$ and $\mathcal{E}_R$ by $x$, and the orthogonal direction in the plane spanned by $\Omega_L$ and $\Omega_R$ by $y$. Then the first phase-matching condition becomes
\begin{align}
k_{1L,x} + k_{2L,x} = k_{1R,x}  + k_{2R,x}, \\
\label{matchingy}
k_{2L,y} = k_{2R,y},
\end{align}
where $\lvert k_{i\alpha} \rvert = \sqrt{k_{i \alpha,x} +k_{i \alpha,y} }$, $i = 1, 2$, $\alpha = L,R$.
With this coplanar scheme, we set $k_{1L, x} = - k_{2L, x}$ and $k_{1R, x} = - k_{1R, x}$. If none of the momenta had $y$ components, these constraints on the $x$ components would clash with Eq.~\eqref{matchingoftheenergies}; Eq.~\eqref{matchingy} allows all of the phase matching conditions to be simultaneously satisfied. Experimentally, this means the need for a coplanar laser scheme.  

\section{Protocol}
\label{sec:preparation}

In this section, we investigate an experimental protocol for implementing an interacting one-dimensional system using Rydberg stationary polaritons. The idea is to initially produce a stationary pulse of light that closely resembles the ground state of a non-interacting one-dimensional model, and tune the interaction to achieve non-trivial correlations.

Initially, atoms are trapped and cooled in a magneto-optical trap. The atoms are then transfered to a cigar-shaped trap, so as to freeze the transverse degrees of freedom of the system and ensure that an effective one-dimensional description is applicable. This can be done either by adiabatically turning on an Ioffe-Pritchard trap, which transforms the  trapping potential into a one-dimensional confining potential \cite{Pritchard:1983dd}, or by adiabatically loading a one-dimensional far-off-resonant-optical dipole trap  
\cite{GRIMM200095}.

\begin{figure}[t]
\includegraphics[width=\columnwidth]{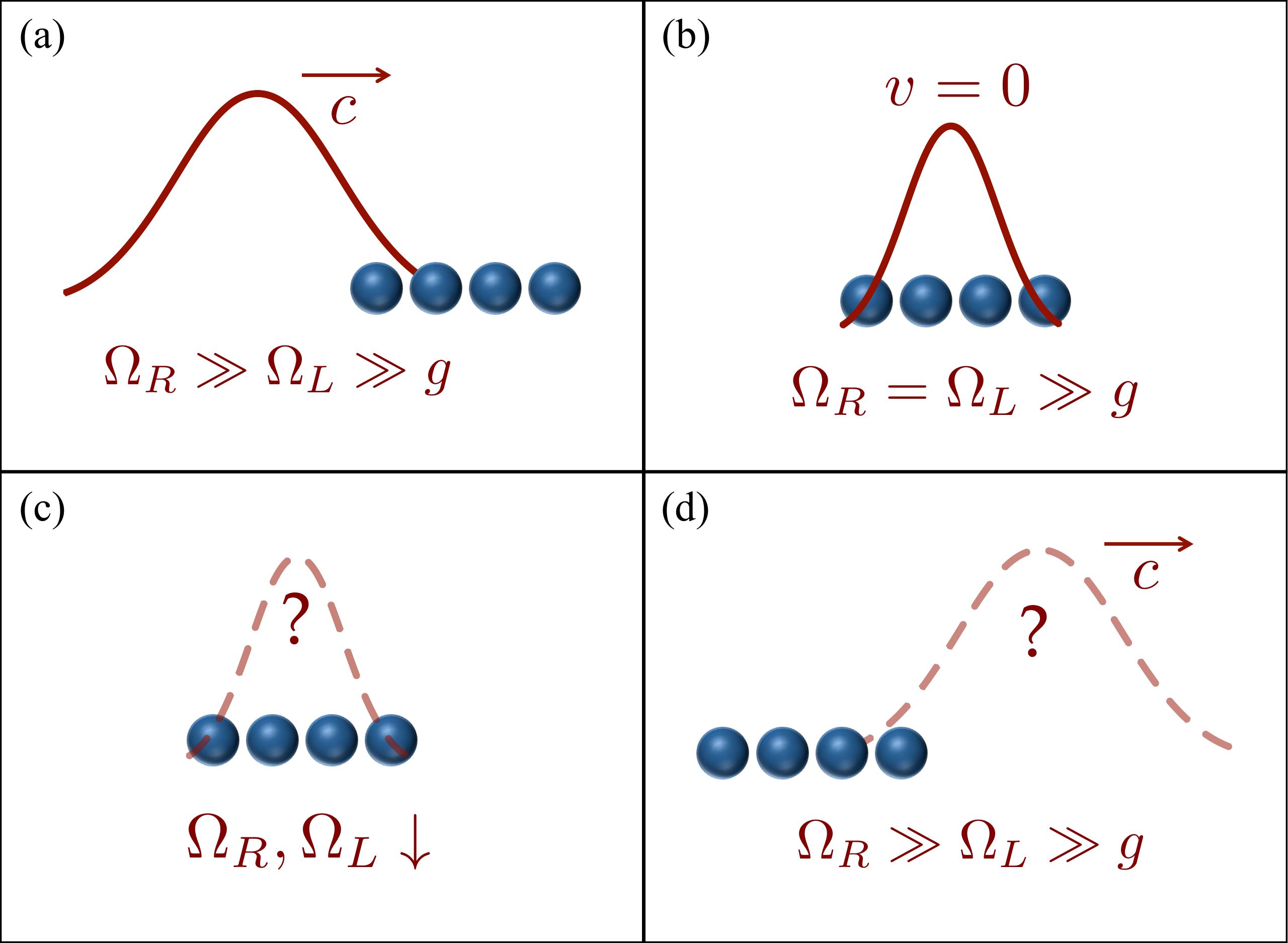}
\caption[scheme]{\label{scheme} Four major steps of the protocol for
  generating and observing non-trivial one-dimensional behavior with
  stationary polaritons discussed in Sec.~\eqref{sec:preparation}. (a)
  A pulse $\mathcal{E}_R$ propagates towards a medium of atoms in a
  cigar-shaped trap, overlapped by two coplanar lasers of Rabi frequencies
  $\Omega_R$ and $\Omega_L$. Initially, $\Omega_R \gg \Omega_L$, and the physics
  is that of the standard EIT scheme. (b) By tuning the Rabi
  frequencies so as to make $\Omega_R = \Omega_L$, one reaches the
  condition of stationarity. (c) Both Rabi frequencies are now
  decreased, while keeping the stationarity condition, so as to change
  the effective one-dimensional scattering length, and to generate
  non-trivial correlations in the pulse. (d) The Rabi frequency
  $\Omega_R$ is quickly increased so as to retrieve the photonic pulse
  while preserving the many-body correlations achieved in the previous stage.}
\end{figure}

Next, a pulse of right-propagating photons associated with the field $\mathcal{E}_R$ of the previous section is prepared. This pulse is focused onto the trapped cold atoms along the longitudinal axis of the trap, with the Rayleigh range of the focus chosen to match the atomic cloud size, which guarantees a homogeneous Rabi frequency. The control lasers are also on, and focused less tightly onto the atom cloud, so that both counter-propagating control fields are well overlapped with the spatial mode of $\mathcal{E}_R$. Initially $\Omega_R \gg \Omega_L$, and $\Omega = \sqrt{\Omega^2_L + \Omega^2_R} \gg g$. The dynamics is then initially that of the standard EIT scheme. As the photons enter the medium, they remain mostly photonic. The beginning of this step is depicted in Fig.~\ref{scheme} (a).

The pulse $\mathcal{E}_R$ is compressed as it enters the medium and propagates slowly and without dissipation as a primarily photonic dark polariton. Once the pulse has fully entered the medium, we decrease $\Omega_{R}$ and increase $\Omega_{L}$, while keeping $\Omega \gg g$, until the stationarity condition $\Omega_L = \Omega_R$ is reached. The dark polariton speed, given in Eq.~\eqref{speedofstationarypolariton}, goes to zero, and the initial pulse is converted to a  stationary pulse of dark polaritons. The condition $\Omega \gg g$ ensures that the polaritons remain mostly non-interacting. This stage is depicted in Fig.~\ref{scheme} (b). At this point, the pulse spectrum should be dominated by the mode $k = 0$, which is the ground state associated with the dark polariton quadratic dispersion $\displaystyle\frac{k^2}{2m}$. The EIT window naturally ensures that only components around $k = 0$ are not dissipated.

The next step is to adiabatically turn on the interaction by reducing
$\Omega$ relative to $g$, while preserving the stationarity condition
$\Omega_L = \Omega_R$. The dark polaritons start to interact as their
atomic component increases. The interaction will lead to a non-trivial
state for the photonic state. In the dilute regime of polaritons, we
tune $\Omega$ adiabatically so as to get a zero one-dimensional
scattering length, similarly to what is done with magnetic fields in
the context of Feshbach resonances \cite{Bienias:2014bc}, leading to a
change in the polaritons' interaction strength, and their effective
scattering length. In other words, the effective scattering length can
be adjusted (independently of the group velocity) by tuning the ratio
of atomic and photonic components in the dark polariton. Depending on
the value of the effective scattering length, a few different sorts of
one-dimensional physics can be realized. For example, $a_{1\text{D}} < 0$
corresponds to the Lieb-Liniger model discussed in
Sec.~\ref{sec:1dsystems}; $a_{1\text{D}} \to 0^-$ realizes the
Tonks-Girardeau gas; and $a_{1\text{D}} \to 0^+$ realizes yet another
regime not discussed here, denoted as the super Tonks-Girardeau gas \cite{Astrakharchik:2005fz}. This step is depicted in Fig.~\ref{scheme} (c).

Currently, we cannot predict the exact $\Omega$ for which each of
these regimes is realized; this is to be inferred from the decay and oscillation rates in the density-density correlation function. In the context of standard EIT polaritons, it has been suggested that the regime where the scattering length approaches $0$ may be found when $\Omega \sim \pm \Delta$ \cite{Bienias:2014bc}. On one hand, the required conditions in a stationary EIT scheme might be similar, since, at least on two-photon resonance, this scheme is very closely related to the standard EIT one. However, there are a few differences that might change the effective interaction. For one, the dark and bright polariton masses and dispersion relations are different. Moreover, there are five, as opposed to three, polariton branches, which could be important in determining the effective interaction. A calculation of the scattering length for stationary dark polaritons is, however, beyond the scope of this work. 

The last step of the protocol is the pulse retrieval. At this
point, we increase $\Omega_R$ relative to $\Omega_L$ and $g$, so as to
convert the polaritons to right-moving photons, as depicted in Fig.~\ref{scheme} (d). This must be done
quickly, since a slow passage would simply revert the non-trivial
state back to its original form. One can then probe the
density-density correlation and its oscillation and decay pattern to
investigate the physics of the state generated by the protocol.

\section{Conclusions}
\label{sec:conclusions}
Owing to the interaction properties of Rydberg atoms, there is strong interest in Rydberg dark polaritons as a platform for enabling photon-photon interactions and, in particular, many-body simulations. In this work, we have investigated the implementation of effective one-dimensional many-body models with Rydberg dark polaritons through a stationary-light scheme. The idea can be summarized in the following steps: creating a stationary-light pulse on top of a one-dimensional atomic gas; tuning the lasers used in achieving the stationary pattern to control the polaritons' interaction strength; and retrieving the pulse in order to probe the non-trivial state generated by the interaction. 

A successful implementation of one-dimensional physics would manifest
in photonic intensity-intensity correlations. For example, in the case
of effective repulsion between polaritons, we expect the polaritons to
fermionize as the effective one-dimensional scattering length
approaches $0$, the correlations then resembling those of non-interacting
fermions. In the manuscript, we discussed the correlation function for the ground state. In practice, obtaining the correlations can be a bit more involved, since we have not considered, for example, effects of the longitudinal confinement. This could be accounted for quantitatively by adding a longitudinal trapping potential to the one-dimensional Hamiltonians discussed in this work. A quantitative comparison is also dependent on the initial pulse shape (or the photon density, which will reflect in the polariton density), which introduces an extra degree of complication. Nevertheless, at a qualitative level, oscillations characteristic of fermionization should persist.

Our work sets a series of milestones to be experimentally achieved so
that Rydberg polaritons may be used as a reliable platform in the
simulation of many-body systems. The first milestone would be the
preparation of a stationary light pulse on top of a cold gas, followed
by its retrieval. In the absence of interactions, experimental
observations of the retrieved pulse should, within a certain degree of precision, coincide with those of the incident pulse. Once this process has been mastered, the next benchmark %, now more specifically in the context of one-dimensional systems, 
 would be the generation and retrieval of stationary light pulses in the context of a one-dimensional atomic gas. Finally, the last extension would be to study the effects of the interaction. 

Our investigation has exposed a few questions that should be the subject of further work. One such question is whether the EIT frequency window for stationary pulses is the same as for standard EIT. A broader EIT window allows for narrower spatial pulses. As discussed in the manuscript, the Hamiltonian responsible for the stationary EIT can be split into symmetric and anti-symmetric sectors. While the symmetric sector physics is the same as that of EIT, the anti-symmetric sector is akin to a two-level system, leading to dissipation. We found that, away from two-photon resonance, these sectors are coupled, which could lead to a narrowing of the stationary EIT window. 

The standard EIT window is usually determined in a semi-classical theory, with atoms treated quantum-mechanically and light as a classical field. Since the first proposals for stationary light were made at a fully-quantum level (with light as a quantum field), there has not been, to our knowledge, studies of the EIT window for stationary light. 
Our preliminary investigation has faced the challenge that the semi-classical approach is missing some features of the fully-quantum Hamiltonian, namely the coupling between symmetric and anti-symmetric sectors. 
Understanding this matter more deeply is important, since a narrowing of the EIT window may impose further limitations on stationary-light schemes than those discussed here.

Notwithstanding these caveats, realizing the proposal outlined above will mark significant progress in the fields of many-body simulations and Rydberg physics.

\begin{acknowledgments}
The authors would like to thank R. G. Pereira and A. Steinberg for helpful
discussions. This work was supported by the Natural Sciences and Engineering Research Council of Canada (NSERC).
\end{acknowledgments}

\bibliography{bibliography.bib}

\end{document}